# Low-dimensional NbO structures on the Nb(110) surface: scanning tunneling microscopy, electron spectroscopy and diffraction

**A.S. Razinkin, E.V. Shalaeva and M.V. Kuznetsov**

*Institute of Solid State Chemistry, Ural Branch, Russian Academy of Sciences,
620041, GSP-145, Yekaterinburg, Russia*

**Abstract**

X-ray photoelectron spectroscopy and diffraction (XPS, XPD) and scanning tunneling microscopy (STM) have been used for study of $NbO_x$-structures on the Nb(110) surface. It is shown that niobium atoms are ordered to form a two-dimensional superstructure with equidistant spacing between the chains of niobium atoms. Chemical shifts of Nb3*d*- and O1*s*-levels demonstrate that the oxide layer corresponds to niobium monoxide NbO and the most part of oxygen in chemisorbed state is localized on the surface. A model of quasi-ordered structure of NbO on the Nb(110) surface is proposed.

*Corresponding author*: e-mail kuznetsov@ihim.uran.ru,
Phone: +7 343 362 33 56, Fax: +7 343 374 44 95

Self-assembling structures on the surfaces of solid materials (substrates) are an important component of the advanced technologies of nanoelectronics and spintronics. A comprehensive study of such low-dimensional objects is necessary for understanding their formation mechanisms and useful physical and chemical properties. This paper presents the results of investigation into the low-dimensional oxygen-induced surface structures on the Nb(110) surface. The authors have been tasked with establishing the atomic structure of the NbO surface layer, determining the localization of oxygen atoms, and studying the chemical nature of oxide nanostructures on the surface [1-5].

In order to thoroughly investigate the structure and properties of low-dimensional surface objects, it is instrumental to use a number of experimental and theoretical techniques that help extract information on the chemical composition and electron and atomic structures of nanoscale objects. In this study, within one high-vacuum complex (Fig. 1), we have combined scanning tunneling microscopy (STM), X-ray photoelectron spectroscopy (XPS), and X-ray photoelectron diffraction (XPD) with the resolution of chemical states of elements. In the XPS method, photoelectrons emitted from the surface under the effect of $hv$-quanta of X-ray radiation provide information about the concentration of atoms on the surface and about their chemical (valent) state. The XPD technique provides analysis of not only the energy of photoelectrons but also the direction of their escape from the surface, and the diffraction pattern is formed at the expense of the photoelectron scattering on the nearest neighborhood of the emitting atom. Theoretical analysis of diffraction patterns yields unique information about the surface structure. Scanning tunneling microscopy determines structural positions of atoms on the surface.

Low-dimensional structures of NbO-type on a Nb(110) single-crystal face were prepared *in situ* in the high-vacuum system of an electron spectrometer ESCALAB MK II by $Ar^+$-ion cleaning followed by annealing of the Nb crystal at temperatures above 2000 K. As this occurred, ordered NbO structures were formed on the surface due to segregation of oxygen atoms dissolved in the crystal bulk and due to desorption of some Nb-O molecules from the surface into vacuum.

Investigation of the topology of the NbO/Nb(110) surface layer by STM has demonstrated that niobium atoms are ordered to form a two-dimensional superstructure with equidistant spacing between the chains of niobium atoms (Fig. 2. a, b). The superstructure lattice parameters were found to be $a = 1.27$ nm and $b = 3.47$ nm. On the surface of single crystal, the NbO superstructure is oriented in two different directions, with these directions being equally probable. Niobium atoms are positioned as chains along the <110> direction on the NbO(111) face and are oriented relative to the Nb(110) substrate, with the <110> NbO(111) direction coinciding with the <111> Nb(110) direction. The angle between the niobium atomic chains in the different orientations is about 60°.

It has been found for the first time that in the surface structure of NbO/Nb(110), in the XPS spectra, the O1$s$-band has such an appearance that one can suggest the existence of two non-equivalent chemical states of oxygen. That corresponds to two types of oxygen positions on the metal surface, with different nature of niobium-oxygen chemical bonding (Fig.3). The spectra of Nb3$d$ demonstrate an increase in the maximum intensity corresponding to NbO-states ($E_b$ = 203.75 eV) as the photoemission angle decreases down to 15˚. That evidences for the surface nature of the structure thus formed. Analysis of composition and chemical shifts of Nb3$d$ levels demonstrated that the oxide layer corresponds to niobium monoxide NbO. The spectra of valence band of the nanoscale NbO-clusters are close in their structure to the valence band of NbO (they are not shown here).

The resulted STM images do not provide data on the oxygen atoms adsorbed on the surface neither do they indicate localization of oxygen atoms in the surface structure of NbO. Localization of structural positions of the two O-states can be obtained from the XPD experiment of O1$s$-photoelectron scattering on the nearest atoms and from the modeling of photoelectron diffraction in the *SSC*-approximation.

Therefore, the combined STM, XPS and XPD experiments made it possible to describe in detail the structure and the chemical nature of the ordered NbO-clusters on the Nb(110) face. A model of the NbO/Nb(110) surface has been built, where oxygen atoms are localized in the structure of linear NbO-clusters (the state of $O_{II}$ in the spectrum of O1$s$) and as chemisorbed atoms $O_I$ near and between the NbO-clusters. It has been shown that, from the viewpoint of stoichiometry and chemical nature (the bonding strengths of XPS Nb3$d$, O1$s$ and spectra of Nb4$d$-O2$p$-valence states) the surface adsorption structures NbO on Nb(110) correspond to niobium monoxide NbO. The XPD analysis of Nb3$d$- and O1$s$- photoemission with the resolution of chemical states has made it possible to determine the orientation of the surface oxide as NbO(111). A model of quasi-ordered structure of NbO on the Nb(110) surface is proposed.

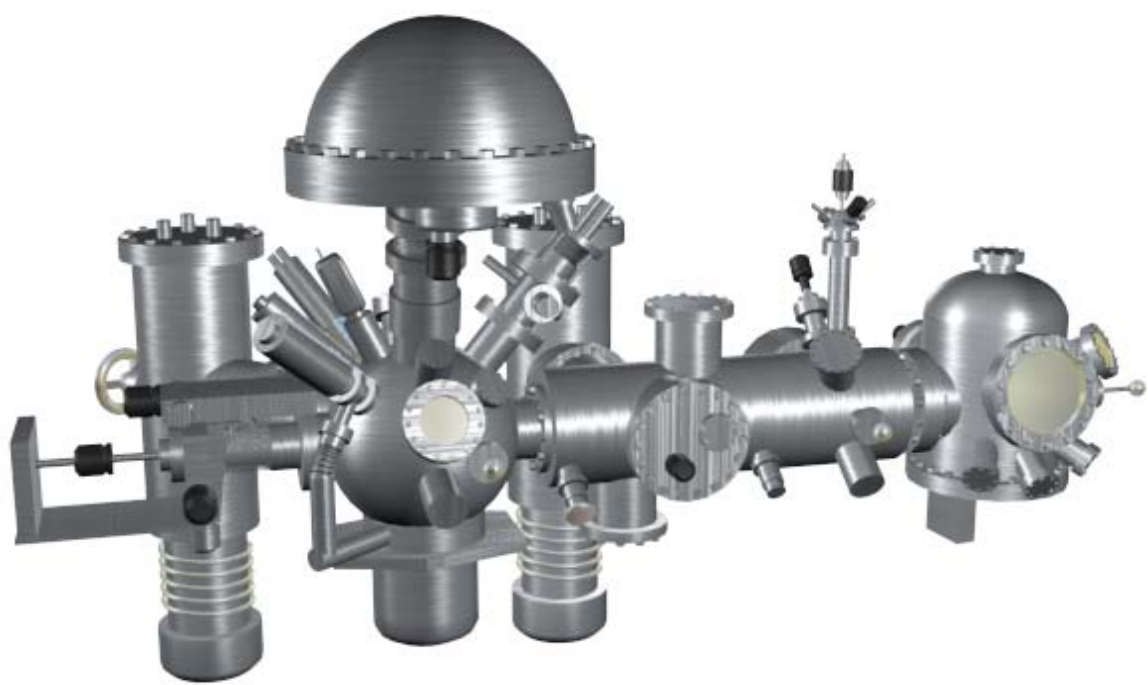

**Figure1**. Electron spectrometer VG ESCALAB MK II + VT STM OMICRON

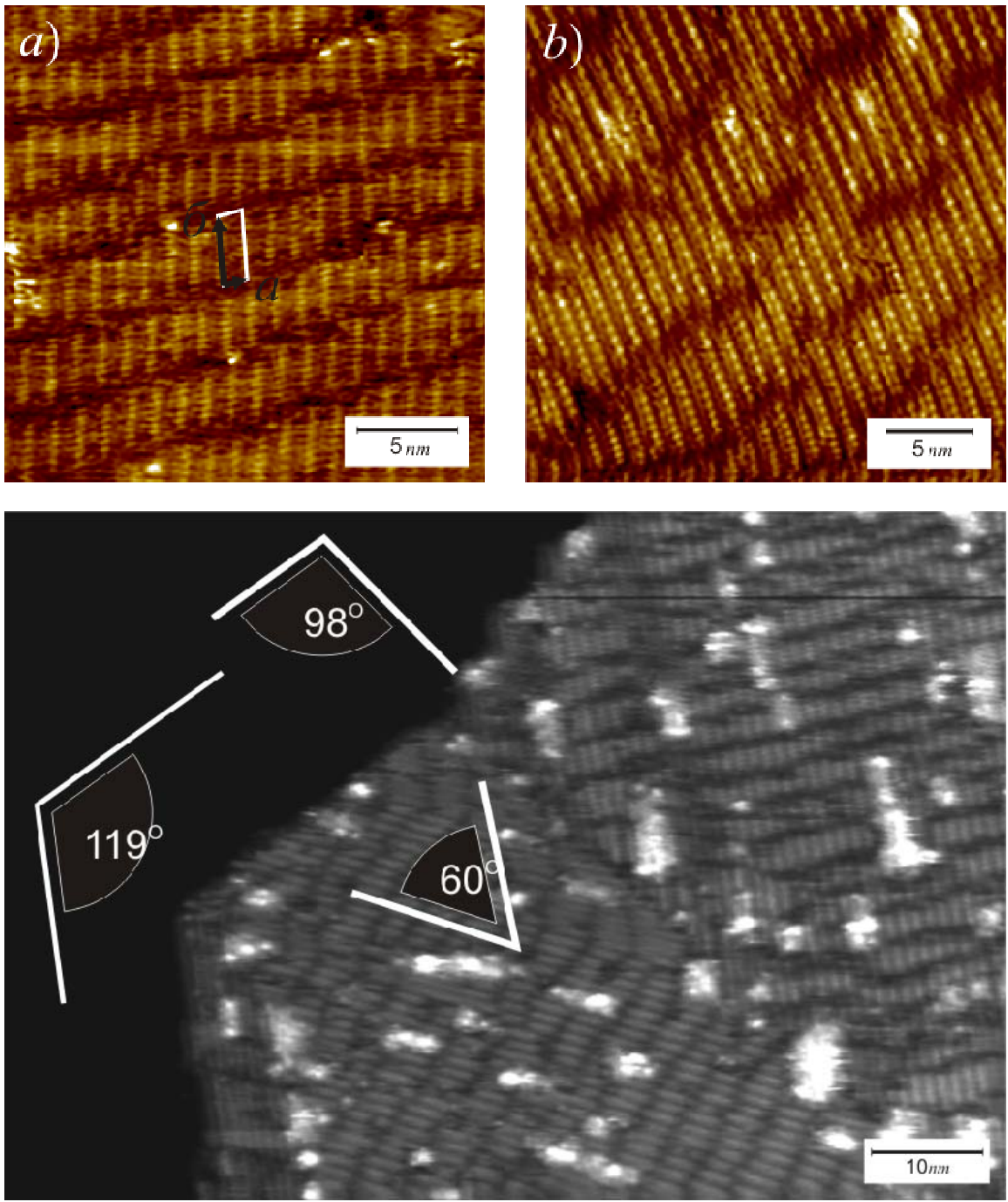

**Figure 2.** STM image of the ordered low-dimensional NbO structure on Nb(110) face ($V$ = 0.18 V, $I$ = 1.5 nA): *a*) Superstructural ordering of niobium atomic chains; *b*) No superstructural ordering of niobium atomic chains; *c*) Mutual orientation of two domains with superstructures of niobium atoms

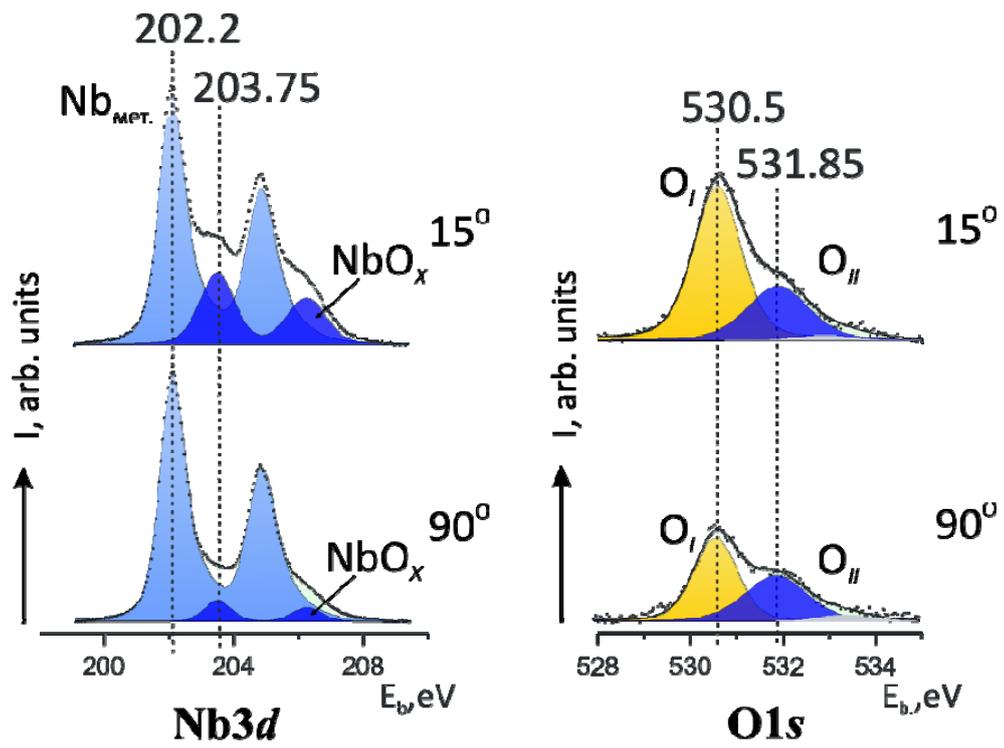

**Figure 3.** Photoelectron spectra of Nb3*d* and O1*s* of the NbO$_x$/Nb(110) surface at photoemission angles 15˚ (*a*) and 90˚ (b) relative to the crystal surface

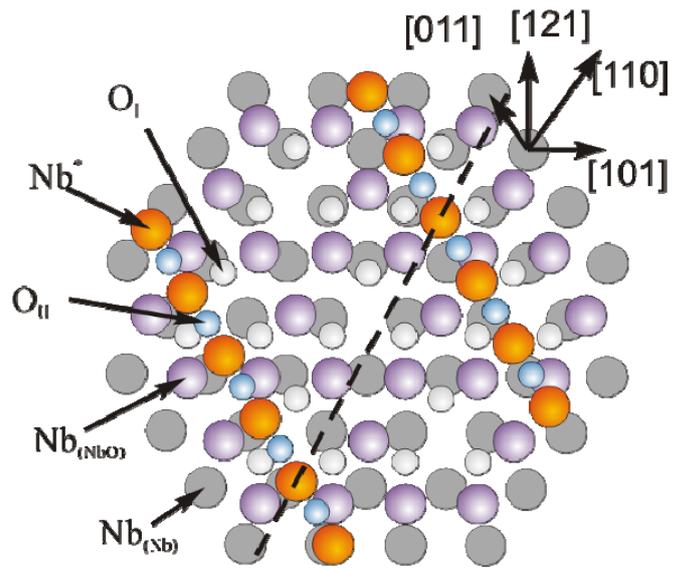

**Figure 4.** A model of the NbO/Nb(110) surface